\documentclass[prl,twocolumn,superscriptaddress,showpacs,floatfix]{revtex4-2}

\usepackage[utf8]{inputenc}
\usepackage{soul}
\usepackage{graphicx}
\usepackage{dcolumn}
\usepackage{braket}
\usepackage{bm}
\usepackage{amsfonts} 
\usepackage{amsmath} 
\usepackage[bookmarksnumbered,pdfpagelabels=true,plainpages=false,colorlinks=true,linkcolor=blue,citecolor=blue,urlcolor=blue]{hyperref}
\usepackage[bookmarksnumbered,pdfpagelabels=true,plainpages=false,colorlinks=true,linkcolor=blue,citecolor=blue,urlcolor=blue]{hyperref}

\def \usach {Departamento de F\'isica, CEDENNA, Universidad de Santiago de Chile, 9170124, Santiago, Chile.}
\def \fcfm {Departamento de F{\'i}sica, Facultad de Ciencias Físicas y Matemáticas, CEDENNA,  Universidad de Chile , Santiago, Chile.}

\def \cbr {Hitachi Cambridge Laboratory, J. J. Thomson Avenue, Cambridge CB3 0HE, United Kingdom}

\begin{document}

\title{A Topological Magnonic Black-White Hole Crystal}

\author{David Galvez-Poblete}
\affiliation{\usach}

\author{Rubén M. Otxoa}
\affiliation{\cbr}

\author{Alvaro S. Nunez}
\affiliation{\fcfm}

\author{Sebastian Allende}
\affiliation{\usach}

\begin{abstract}

The low-energy dynamics of antiferromagnetic spin waves map onto a massive Klein-Gordon scalar field, providing a solid-state platform for analogue-horizon physics. We show that an inhomogeneous spin current generates an effective magnon flow capable of forming black- and white-hole horizons. Horizon-pair cavities exhibit resonant transport and superradiant-like amplification mediated by magnon-antimagnon mixing. Our central result is that a periodic arrangement of submagnonic and supermagnonic regions forms a non-reciprocal magnonic crystal in which entering the supermagnonic regime makes negative-norm propagating channels available and enables the bulk gap to close and reopen. This drives a topological transition characterized by a change in the Zak phase and the emergence of a hybrid magnon-antimagnon edge state. These results establish a direct connection between analogue-horizon physics and topological magnonics, providing an electrically tunable route for controlling magnonic band topology.
\end{abstract}

\maketitle

\textit{Introduction.}--- Spintronics has traditionally relied on ferromagnets to exploit the spin degree of freedom as an information carrier \cite{baltz2018antiferromagnetic}. However, magnons provide a versatile platform for transport phenomena beyond conventional electronics, as their propagation can be controlled by external magnetic fields, spin currents, and geometrical confinement \cite{Lenk2011}.

Recently, antiferromagnetic spintronics has emerged as a promising alternative, driven by its ultrafast terahertz (THz) dynamics, absence of stray fields, and inherent relativistic-like behavior \cite{baltz2018antiferromagnetic}. Crucially, the low-energy dynamics of antiferromagnetic spin waves can be mapped onto a massive Klein--Gordon equation \cite{tatara2020magnon}, establishing a natural connection between condensed matter systems and concepts from relativistic field theory.

This connection becomes especially relevant in the context of analogue gravity, first introduced by Unruh in 1981 \cite{unruh1981, hawking1974}. Analogue gravity demonstrates that the kinematic structure of curved spacetime and event horizons can be emulated in laboratory setups \cite{Visser1998,Barcel2011}. Since then, a wide variety of platforms, including fluids \cite{Rousseaux2008}, superfluids \cite{Volovik2001}, optical systems \cite{Philbin2008,Belgiorno2010}, and Bose-Einstein condensates \cite{Barcel2001, Garay2000}, have been proposed to reproduce phenomena such as black-hole and white-hole horizons, Hawking radiation, and superradiance in experimentally accessible settings. In magnetic systems, magnonic black-hole analogues have been theoretically established \cite{roldan2017magnonic, bassant2025thermal, gnoatto2025controlling, Doornenbal2019}. In the presence of spin currents, magnons experience a Doppler shift that effectively induces a flow velocity. When this flow becomes spatially inhomogeneous, it can give rise to regions where counter-propagating modes are suppressed, defining magnonic horizons analogous to black-hole and white-hole configurations \textcolor{blue}{\cite{roldan2017magnonic, bassant2025thermal, gnoatto2025controlling, negative_energy_stability}}.

Despite these conceptual advances, predicting measurable superradiant amplification and topological transitions via analogue gravity remains a profound challenge. In this work, we demonstrate that a spin-current-driven antiferromagnet can realize a magnonic black-hole-white-hole crystal, where propagating magnons experience a periodic array of horizon-like interfaces. We demonstrate that such configurations support distinct transport regimes, including resonant cavity behavior and superradiant-like amplification mediated by magnon-antimagnon mixing \textcolor{blue}{\cite{kamra2023antimagnonics, vool2026superradiant,Errani2025}}. Furthermore, we establish that the periodic arrangement of these horizons naturally hosts asymmetric band structures exhibiting a flow-controlled topological transition characterized by a change in the Zak phase \cite{magnon_topology_zak}. By proposing realistic implementations in standard antiferromagnets like MnPSe$_3$, driven by achievable spin-current densities, our results elevate magnonic horizons from theoretical curiosities to viable, electrically tunable platforms for topological transport and superradiance.

\textit{Theoretical Model.}--- 
To establish the theoretical framework for these phenomena, we consider an antiferromagnetic material under the action of an adiabatic spin-transfer torque (STT) induced by an applied charge current \cite{Fujimoto2021}, whose magnonic dynamics is described by the following equation (see the Supplementary Material (SM) for a detailed derivation):
\begin{equation}
    -K_z \psi + A \frac{\partial ^2}{\partial x^2}\psi - \frac{1}{\Lambda} \left(\frac{M_s}{|\gamma|}\right)^2 \left(  \partial_t + u \partial_x \right)^2 \psi =0 
\end{equation}
Here, the complex field $\psi$ represents small transverse deviations of the N\'{e}el magnetization. The parameter $M_s$ is the saturation magnetization, $\gamma$ is the gyromagnetic ratio, and $\Lambda$ characterizes the strength of the exchange interaction. The constant $A$ corresponds to the exchange stiffness and $K_z$ is the uniaxial anisotropy parameter that favors alignment of the N\'{e}el vector along the $z$-axis. Finally, $u \equiv P j_c/e M_s$ is the effective flow velocity, proportional to the applied current density $j_c$ and $P$ is a dimensionless coefficient characterizing the efficiency of the adiabatic spin-transfer torque.
From the previous expression, we can identify the massive scalar Klein-Gordon equation:
 \begin{equation}
    (\partial_t+u(x) \partial_x)^2 \psi - v^2 \partial_x^2\psi + m^2(x) \psi =0,
\end{equation}
 where we have defined the parameters $v \equiv \frac{|\gamma|}{M_s}\sqrt{A\Lambda}$, which plays the role of the spin wave velocity, and $\quad m \equiv \frac{|\gamma|}{M_s} \sqrt{K_z \Lambda}$, which defines the effective mass (gap) of the field. This equation constitutes the cornerstone of our analysis. A natural length scale of the system is given by $a \equiv \sqrt{A/K_z}$.


\begin{figure}
    \centering
    \includegraphics[width=1.0\linewidth,trim=4cm 3cm 10cm 0cm, clip]{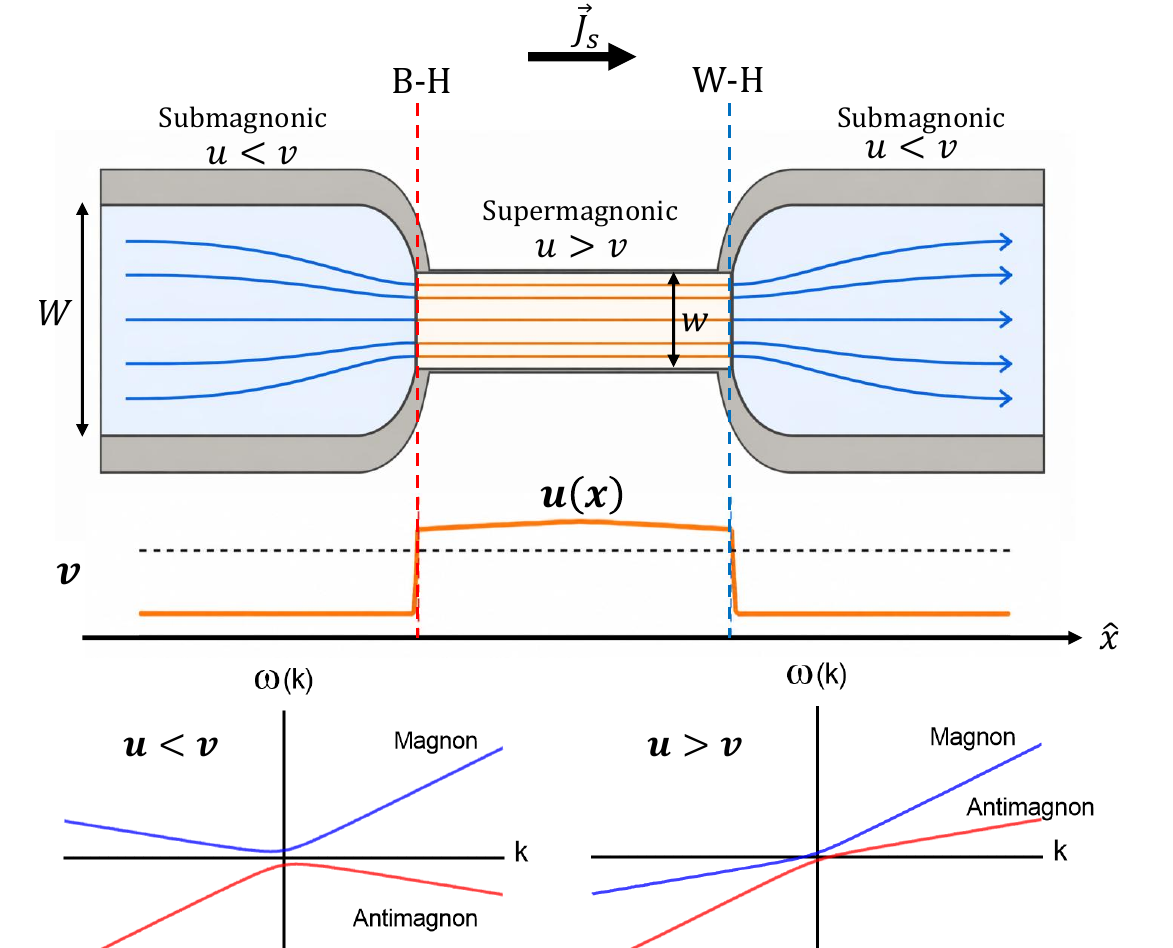}
    \caption{Schematic setup for black-hole (B-H) and white-hole (W-H) horizons in an antiferromagnetic system induced by a geometrical constriction, $w \ll W$ , and an applied spin current. The central region hosts a supermagnonic regime ($u>v$), whereas the outer regions remain submagnonic ($u<v$). The lower panel displays the spatial profile of the effective flow velocity $u(x)$, with the horizons located at the transition between both regimes.}
    \label{fig:placeholder}
\end{figure}

Building upon this relativistic description, we first analyze the formation of a single magnonic horizon. Let us consider an uniform flow $u(x) = u$. We then perform a plane waves solution of the form $\psi = \phi_o \text{e}^{i(kx-\omega t)}$, obtaining the following expression: 
\begin{equation}
    (\omega - uk ) ^2 =v^2 k ^2 + m ^2
\end{equation}
The left-hand side of this equation can be associated with a co-moving frame with $\tilde{\omega} = \omega - uk $, and the corresponding co-moving frame group velocity is given by: 
\begin{equation}
     \tilde{v}_g = \frac{d\tilde{\omega}}{dk } = \pm \frac{v^2k}{\sqrt{v^2 k ^2 + m ^2}}.
\end{equation}
Since $\tilde{v}_g(k\to\infty)=\pm v$, $\tilde{v}_g$ is monotonic in $|k|$, it follows that $|\tilde{v}_g|\leq v$. This latter condition implies that $u-v \leq v_g \leq u+v$. Let us consider $u>0$, corresponding to a flux directed along the positive $x$-axis. The minimum value that the group velocity in the co-moving frame can take is $\tilde{v}_{g}^{min} < u-v$. We are interested in determining whether counter-propagating modes, $v_g<0$, can exist in the system. For such modes to appear, the condition $u<v$ must be satisfied, while still respecting the initial assumption $u>0$. Therefore, when $u>v$, counter-propagating modes with $v_g<0$ cannot exist, and all magnonic excitations are forced to propagate in the positive $x$-direction. This allows us to define two distinct regimes. In the submagnonic region, where $u<v$, counter-propagating modes are allowed. In contrast, in the supermagnonic region, where $u>v$, no counter-propagating modes exist and negative-norm propagating modes emerge (antimagnons). The interface between these two regions, defined by the condition $u=v$ plays the role of a magnonic horizon. Depending on the direction of the propagating waves relative to the interface, this horizon can behave either as a magnonic black-horizon (BH) or as a magnonic white-horizon (WH).

A favorable material route is provided by soft van der Waals antiferromagnets such as MnPSe$_3$ \cite{Liao2024,Jana2024}. An estimated magnon velocity of $v \sim 2 \times 10^3 \text{m/s}$, together with a natural  length $
a\sim 15-20$ $\text{nm}$, gives a required current density of the order of $j \sim 10^{13} \text{A/m}^2$, high but within the upper range of current densities used in spin-torque experiments. Therefore, cavity lengths \(L/a\sim1-5\) correspond to $L\sim 15-100$ $\text{nm}$, which are compatible with nanofabricated constrictions. In a realistic implementation, Joule heating may have important consequences \cite{Zink2022}. In particular, it may renormalize the magnetic parameters through their temperature dependence, thereby shifting the analogue horizons and modifying the transition between the submagnonic and supermagnonic regimes. To mitigate these effects, the required current density could be further reduced by engineering the magnetic exchange interactions in $\text{MnPSe}_3$. Since the exchange interactions in van der Waals magnets are highly sensitive to strain and layer engineering \cite{Ren2023, Pei2018}, the long-wavelength magnon velocity of $\text{MnPSe}_3$ may, in principle, be softened through suitable structural tuning from $v \to \eta v$. In this case, the critical current decreases as $j_c \to \eta j_c$, while the corresponding Joule power dissipation density decreases as $\eta^2$. Thus, even a moderate reduction of the magnon velocity would substantially relax the current requirements while suppressing Joule heating quadratically, making the experimental realization considerably more accessible.



Having established the conditions for single horizons, we now focus on cavity-like configurations. The BH–WH and WH–BH cavities are first analyzed as elementary scattering units of the periodic crystal considered below. Their role is to establish the transfer matrices underlying the band structure and topological properties of the periodic system.

\textit{BH-WH pair configuration.}---
We consider a BH-WH configuration consisting of an inner supermagnonic region bounded by two external submagnonic regions, as illustrated in Fig. 1. Magnonic waves are injected from the left, i.e., from $x\to -\infty$. The flow profile is taken to be piecewise constant, with $u_1<v$ for $x<0$, $u_2>v$ for $0<x<L$, and $u_3<v$ for $x>L$. The interfaces at $x=0$ and $x=L$ correspond to the horizon positions, where $u=v$. Solving Eq.(5) for the wave vector $k$, we obtain
    \begin{equation}
        k = \frac{u \omega \pm \sqrt{v^2 \omega^2 + m^2 (u^2 - v^2)}}{(u^2-v^2)}
    \end{equation}

In the outer submagnonic regions, propagating modes exist only for $\omega > \omega_c \equiv m\sqrt{(1-u_1^2/v^2)}$, whereas inside the supermagnonic region the two real solutions propagate in the same direction, one of them belonging to the negative-norm branch.
Therefore, we propose piecewise plane-wave solutions in each region, as detailed in the SM. These solutions must satisfy the continuity conditions obtained by integrating equation (2) across each interface, i.e, the continuity of the field $\phi$ and $\alpha \phi ' - i \omega u \phi$, with $\alpha =  u^2 - v^2$. 

\begin{figure}[t]
    \centering
\includegraphics[width=1.0\linewidth, trim= 1cm 0cm 0.5cm 1cm]{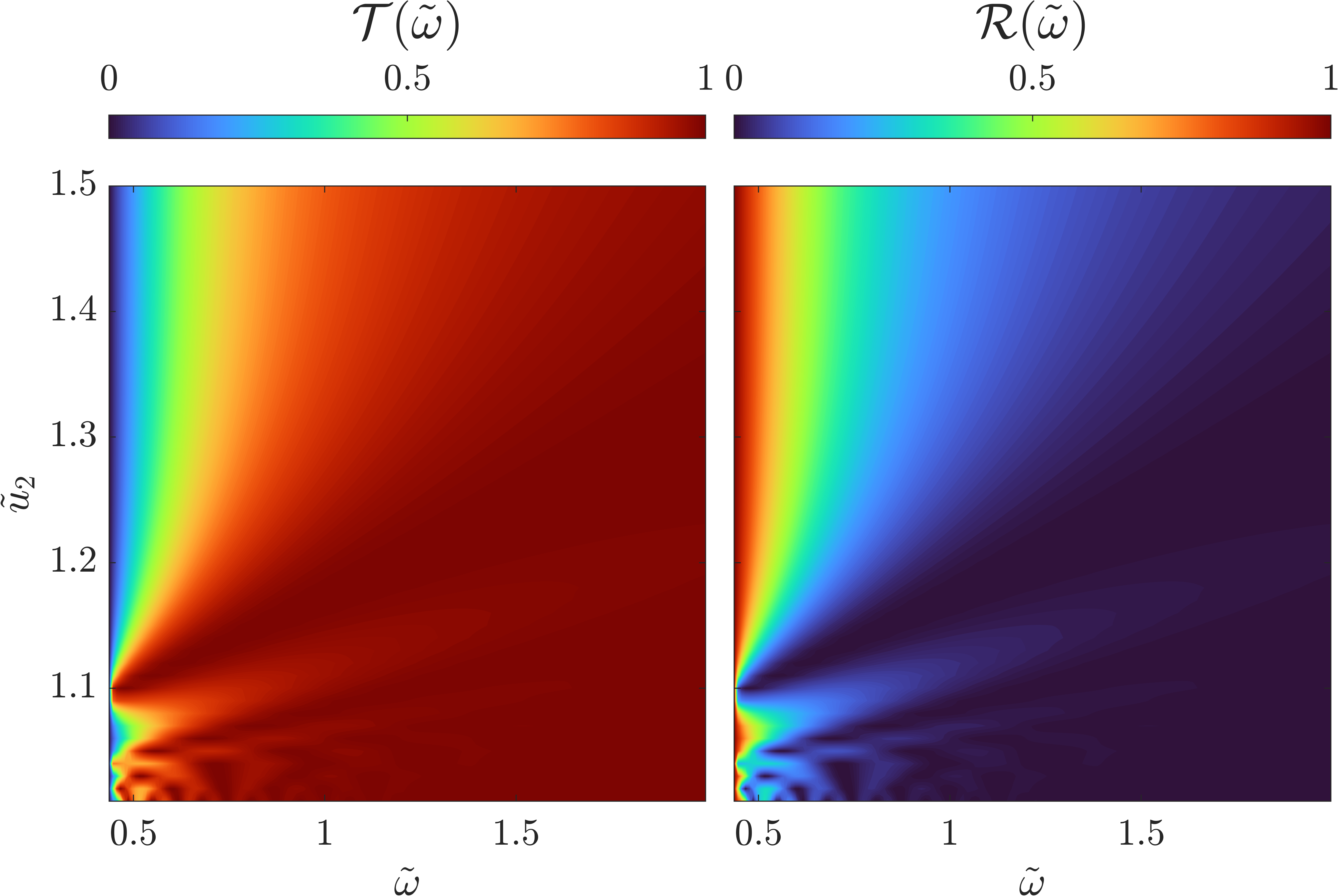}
    \caption{Transmission and reflection coefficient for BH-WH cavity, as a function of dimensionless supermagnonic  velocity $\tilde{u}_2 \equiv u_2 /v$ and frequency $\tilde{\omega} \equiv \omega/m$, for $L = a$ and $\tilde{u}_1 \equiv u_1 /v = u_3/v= 0.9$.}
    \label{fig:placeholder}
\end{figure}

Numerical results for the reflection, $\mathcal{R}$, and transmission, $\mathcal{T}$ coefficients defined in the SM are shown in Fig. 2. As expected, they satisfy the current conservation relation $\mathcal{R}+ \mathcal{T} =1$. Near the critical frequency, the cavity behaves as an effective barrier with suppressed transmission. For suitable values of the supermagnonic velocity and cavity length, resonant transmission emerges due to multiple positive/negative-norm conversions inside the cavity. These resonances become increasingly pronounced for longer cavities and lower supermagnonic velocities, while the system approaches perfect transparency in the high-frequency limit.

\textit{WH-BH pair configuration.}---
We next consider the complementary $WH-BH$ cavity, consisting of an inner submagnonic region surrounded by two supermagnonic regions. Owing to the supermagnonic asymptotic regions, no reflected waves are supported, and the transmitted field contains both positive and negative-norm modes. The corresponding piecewise solutions and matching conditions are summarized in the SM.

The resulting transmission coefficients are shown in Fig. 3. Since the outgoing region supports both positive and negative-norm currents, current conservation takes the form $\mathcal{T}_{+}-\mathcal{T}_{-}=1$, revealing amplification through positive- and negative-norm mode mixing at the horizon. Here, energy is extracted from the STT-induced flow, analogous to the extraction of rotational energy from a rotating black hole in gravitational superradiance. We therefore identify this process as a superradiance-like phenomenon. At low frequencies, where the modes inside the submagnonic cavity become evanescent, positive/negative-norm mixing is strongly enhanced, leading to large values of both transmission coefficients and therefore to an effective amplification. This effect becomes more pronounced for longer cavities. In contrast, at high frequencies the negative-norm contribution is progressively suppressed, yielding the asymptotic behavior $\mathcal{T}_{+} \to 1$, $\mathcal{T}_{-} \to 0$.

\begin{figure}[t]
    \centering
    \includegraphics[width=1.0\linewidth, trim= 1cm 1cm 1.5cm 1cm]{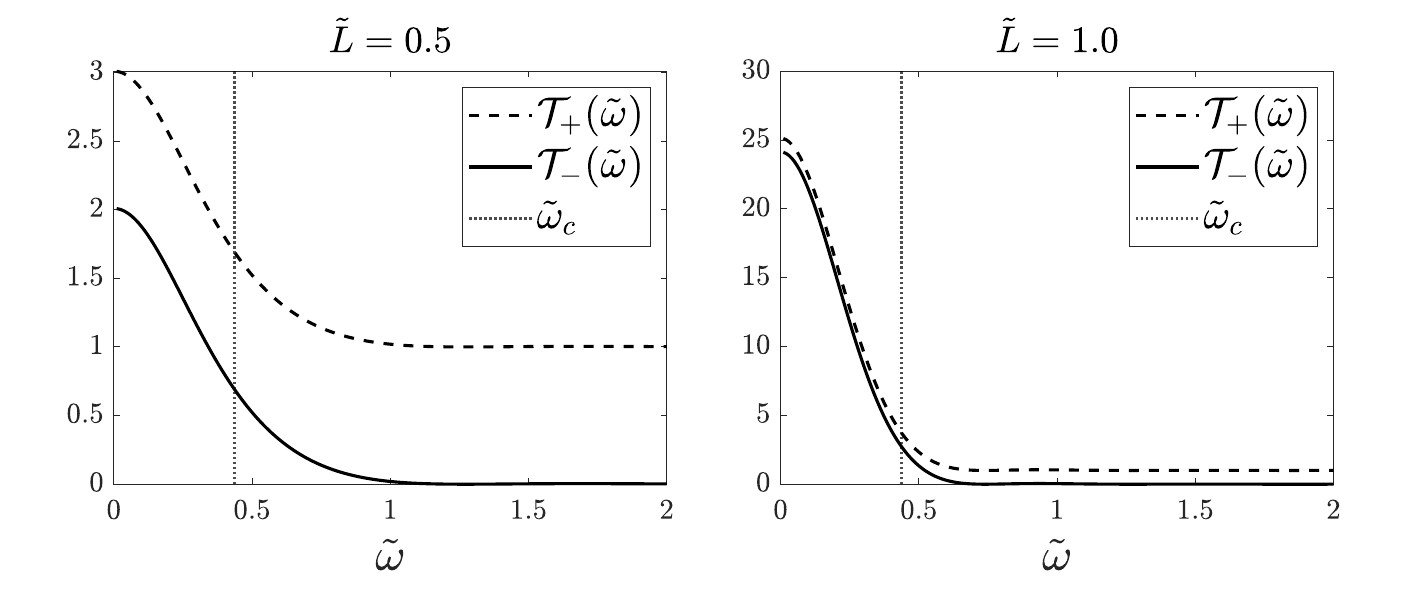}
    \caption{Negative $\mathcal{T_{-}(\tilde{\omega})}$ and positive $\mathcal{T_{+}(\tilde{\omega})}$ transmission coefficients for different cavity $\tilde{L} \equiv L/a $, with $\tilde{u}_2 \equiv  u_2/v = 0.9$ and $\tilde{u}_1 \equiv u_1 /v= u_3/v = 1.1$.In this configuration, the dimensionless critical frequency is given by $\tilde{\omega}_c \equiv \omega_c /m = 0.4359 $.}
    \label{fig:placeholder}
\end{figure}

\textit{Crystal of magnonic Black-White Hole.}---
Extending these concepts beyond isolated structures, We now consider a system consisting of a periodic array of cavities, formed by alternating submagnonic and supermagnonic regions separated by magnonic horizons. For a periodic array, the field amplitudes in two consecutive unit cells are related through a transfer matrix $\mathcal{M}(\omega)$,
\begin{equation}
    \mathbf{A}_{n+1}=\mathcal{M}(\omega)\mathbf{A}_{n} = e^{iqd}\mathbf{A}_{n}
\end{equation}
where, in the last equality, we have imposed the Bloch condition. Therefore, the allowed bands are obtained from the full eigenvalue equation
\begin{equation}
    \det\!\left[\mathcal{M}(\omega)-e^{i\tilde q d}\mathbb{I}\right]=0.
\end{equation}
Since \(\mathcal{M}(\omega)\) is generally complex, the band spectrum is not inferred from the real trace condition alone. Instead, propagating Bloch modes are identified from the real solutions of the full eigenvalue problem. Solving this condition numerically yields the asymmetric band structures shown in Fig.~4.

\begin{figure}
    \centering
    \includegraphics[width=0.9\linewidth]{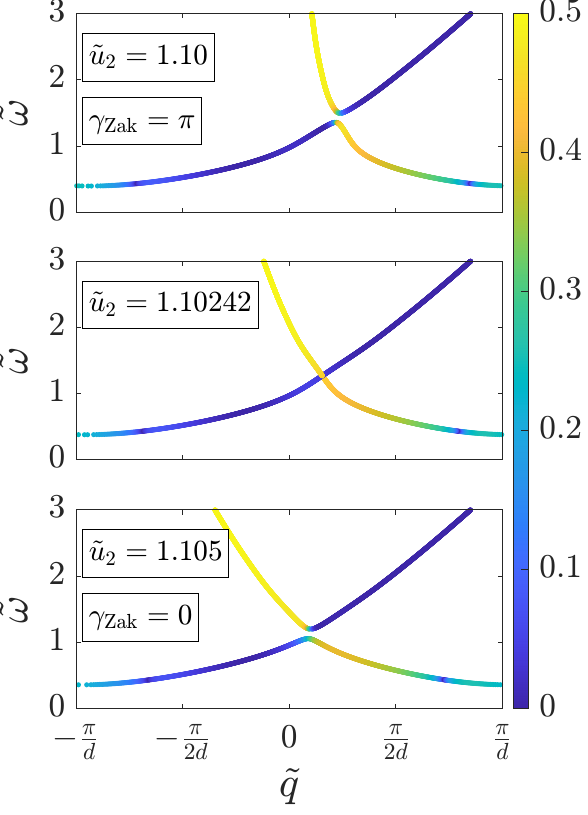}
    \caption{Band structure for different supermagnonic flux, and their topological Zak phase for $\tilde{L}_a = \tilde{L}_b = 1.0$, $\tilde{u}_1 = 0.9$ and $\tilde{d} = \tilde{L}_a + \tilde{L}_b$. The dimensionless wave-number is defined as $\tilde{q} = qa$. The colorbar indicates the relative presence of antimagnonic waves in the band, $f_{AM}$.}
    \label{fig:placeholder}
\end{figure}

The inhomogeneous flow induces a frequency-dependent momentum shift, linear in \(\omega\), which is responsible for the non-reciprocal and asymmetric band structure of the  system (see SM). In addition, the system displays a low-frequency gap that is not determined solely by the critical frequency $\omega_c$, but also depends strongly on the length of each region and on the supermagnonic velocity $u_2$. This implies that, for certain parameter choices, wave propagation can occur even at frequencies below $\omega_c$, although the waves are locally evanescent in the submagnonic regions. This behavior arises from the nontrivial coupling between the different regions, as well as from the interplay between magnonic and antimagnonic branches, which can be observed from the presence of antimagnonic waves within the bands. In particular, we define the relative antimagnonic weight of each Bloch state as:
\begin{equation}
    f_{AM} = \frac{\sum_{\Omega_\nu<0} L_\nu|\Omega_\nu|^2 |c_\nu |^2}{\sum_\nu L_\nu|\Omega_\nu|^2 |c_\nu|^2}
\end{equation}
Here $c_\nu$ denotes the amplitude of each plane-wave component, $\Omega_\nu = \omega - u_jk_\nu$ its comoving frequency, and $L_\nu$ corresponds to the propagation length within one unit cell.

As the supermagnonic velocity $u_2$ is varied, the gap between the bands undergoes a closing and subsequent reopening. The gap-closing condition is given by:

\begin{equation}
    \frac{|k_{\rightarrow} - k_{\leftarrow}|}{2}L_a + \frac{|k_{2,+} - k_{2,-}|}{2} L_b \approx n \pi 
\end{equation}

with $n$ an integer, and $L_a$ and $L_b$ being the lengths of the submagnonic and supermagnonic regions, respectively. In particular, the gap closing shown in Figure 4 corresponds to $n=4$ at the dimensionless frequency of  $\tilde{\omega}_g = 1.271$. Another gap closing is found at the same frequency for $\tilde{u}_2=1.2075$, corresponding to $n=3$. The gap closing associated with $n=2$ occurs at $\tilde{\omega}_g=0.738$. In contrast, the $n=1$ gap closing is not observed, since it would occur at a frequency below the lower band gap. Higher-order gap closings associated with larger values of $n$ are, in principle, possible. However, for the frequency range considered here, only the $n=4$ and $n=3$ conditions are accessible. Gap closings with $n>4$ would require larger frequencies, corresponding to higher-order Bloch phase accumulations across the unit cell.

The previous relation indicates that the gap closes when the accumulated phase over one unit cell satisfies a Bloch phase-matching condition. At the critical configuration where the gap closes, the spectrum displays non-reciprocal Dirac-like points. This gap-closing and reopening process is accompanied by an underlying topological transition, characterized by a Zak phase \cite{Vanderbilt2018}:
\begin{equation}
    \gamma_{Zak} =  \int_{BZ} A(q) dq ;\quad A(q) = i \langle u_{nq} |  \partial_q u_{nq} \rangle
\end{equation}
where $u_{nq}$ represents the periodic part of the Bloch function. In Figure 4, prior to the transition, the lower band is characterized by a Zak phase of $\pi$, signaling a nontrivial topological phase. After the gap reopens, the lower band becomes topologically trivial, with Zak phase 0. The origin of this topological transition is that the system behaves similarly to a dimerized Kronig–Penney lattice \cite{Reshodko2019} or an SSH chain \cite{Su1979}, with the key difference that the transition is controlled by the spin-current-induced flow.

To demonstrate the bulk-boundary correspondence, we analyze a finite chain. In the non-trivial phase ($u_2 = 1.10)$, an edge state emerges inside the bulk band gap, whose localization strongly depends on the termination of the chain. Conversely, in the trivial phase ($u_2 = 1.105)$, the edge mode disappears and no in-gap localized state is found. Importantly, the localized in-gap state is a hybrid magnon–antimagnon state, with an approximate relative antimagnonic weight of ($f_{\rm AM}=0.44$). This demonstrates that the boundary state retains a direct fingerprint of the positive/negative-norm mode structure enabled by the supermagnonic region. 

An important mathematical feature of our model is that the transfer matrix of a unit cell, $\mathcal{M} (\omega)$, can be written as $\tilde{\mathcal{M}}(\omega) = e^{-i \theta(\omega) /2} \mathcal{M}(\omega)$ where the additional phase depends on the spin-current-induced magnonic flow. The auxiliary matrix satisfies $det(\tilde{\mathcal{M}} )=1$, and as shown in the SM, takes the form:
\begin{equation*}
   \tilde{ \mathcal{M}} = \begin{pmatrix}
       C& D \\D^* &C^*
   \end{pmatrix} ; \quad \tilde{\mathcal{M}}^{\dagger} \eta \tilde{\mathcal{M}} = \eta
\end{equation*} 
with $\eta = \sigma_z$ and $C,D$ are complex coefficients that depends on the parameters of the system. Therefore, the auxiliary matrix belongs to the pseudo-unitary group $SU(1,1)$ \cite{Perelomov1977}, whose coefficients satisfies $|C|^2-|D|^2 =1$, same structure that satisfies the transmission coefficients previously studied.  In particular, if $u_{\tilde{q}}$ is an eigenvector of the auxiliary matrix $\tilde{\mathcal{M}}$ (and consequently of the cell matrix $\mathcal{M}$), it follows that $u_{-\tilde{q}} = \sigma_x u_{\tilde{q}}^*$. This implies that $A(-\tilde{q}) = - A(\tilde{q})$. This result, together with the inversion symmetry of the infinite crystal, constrains the Zak phase to the quantized values of $(0,\pi)$.
\begin{figure}
    \centering
    \includegraphics[width=1.0\linewidth, trim= 3cm 8cm 3cm 8cm]{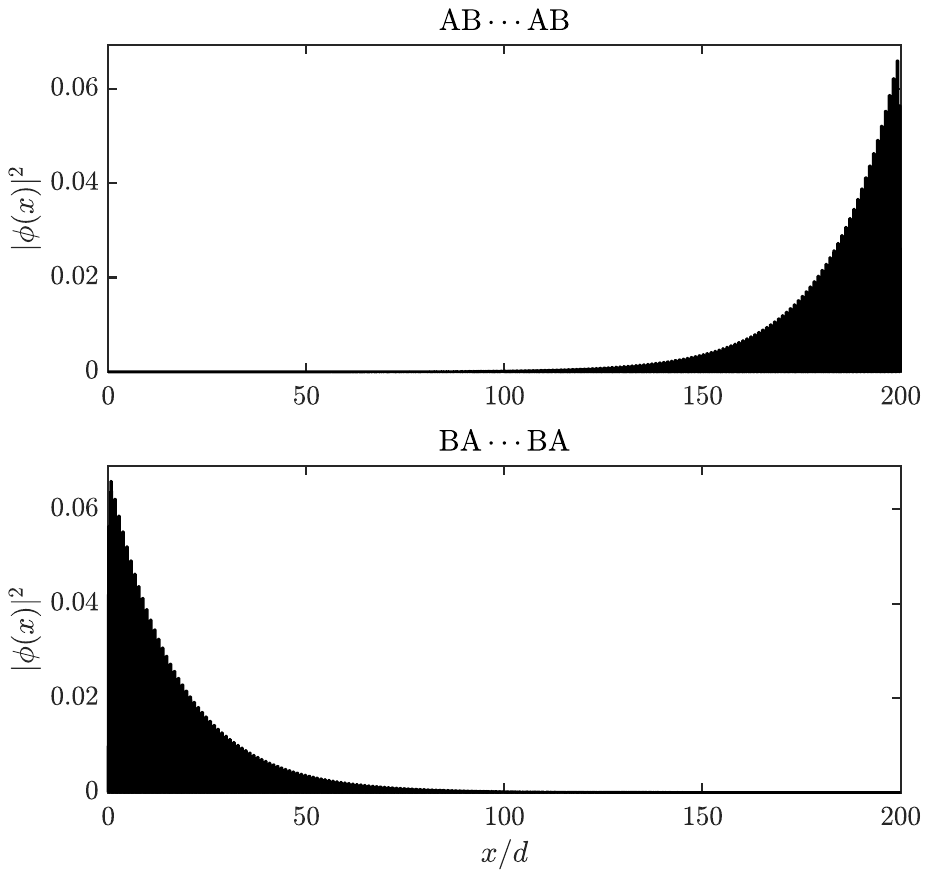}
    \caption{Spatial profile of the edge-state density, $|\phi (x)|^2$, at $\tilde{\omega} = 1.365$ for the topologically nontrivial phase ($\tilde{u}_2 = 1.10$). The upper panel shows the edge state for the finite $AB...AB$ chain configuration, whereas the the lower panel shows the corresponding edge state for the inverted $BA...BA$ chain. Here, $A$ and $B$ denote the submagnonic and supermagnonic regions, respectively. Here, we have defined $d = L_a + L_b $.}
    \label{fig:placeholdera}
\end{figure}
Remarkably, the $SU(1,1)$ structure and the resulting Zak-phase quantization are independent of the existence of supermagnonic regions. Consequently, both the purely submagnonic and the sub-supermagnonic configurations belong to the same symmetry class. The essential difference is that, in the purely submagnonic regime, the gap-closing condition is never satisfied for the two lowest bands, so the bulk gap remains open throughout the parameter space. As a result, the system remains adiabatically connected to a trivial phase with Zak phase 0. In contrast, the emergence of a supermagnonic region enables a gap-closing and reopening process, allowing the Zak phase to change from 0 to $\pi$ and giving rise to topologically protected edge states, see Fig. \ref{fig:placeholdera}. Therefore, analogue horizons do not modify the symmetry class of the system; instead, they provide the physical mechanism that drives the topological phase transition by enabling the bulk gap to close and reopen. 

It is important to note that the edge mode localizes at the interface between the supermagnonic region, denoted by B in the Fig. 5, and the vacuum. This behavior can be understood from the auxiliary matrix $\tilde{\mathcal{M}}$, which determines whether a given frequency belongs to a propagating band or to a bulk gap. In particular, a gap occurs when $|\text{Tr}(\tilde{\mathcal{M}})|>2$. As shown in the SM, the Bloch eigenvectors inside the band gap are necessarily isotropic with respect to the conserved-current (Krein) metric, $v^\dagger \sigma_z v =0$. For the present submagnonic–supermagnonic crystal, this condition requires the simultaneous presence of positive- and negative-norm components, i.e., a hybrid magnon–antimagnon state. Since negative-norm excitations are supported only in the supermagnonic region, the edge mode can only localize at the supermagnonic–vacuum interface.

As a final remark, our analysis has focused on the conservative limit. In realistic insulating antiferromagnets such as $\text{MnPSe}_3$, the intrinsic Gilbert damping \cite{Kamra2018} is expected to be weak, with typical values of the Gilbert parameter in the range of $\alpha \sim 10^{-4}-10^{-3}$ \cite{Tzschaschel2020,Hong2024}. Consequently, dissipation is expected to mainly introduce finite magnon lifetimes without qualitatively modifying the band structure or the topological phase. Likewise, the edge states are expected to acquire finite lifetimes while remaining spatially localized. For the magnonic cavities considered here, dissipation introduces an effective decay rate that becomes significant after approximately $N = 1 /(2 \alpha \tilde{\omega} \tilde{L}) \sim 5 \times 10^2$ round trips inside the cavity. Nevertheless, the mixing of positive and negative-norm modes may also give rise to dynamical instabilities, as reported in related systems. In ferromagnetic systems, the combined action of Gilbert damping and dissipative spin-transfer torques has been shown to stabilize negative-energy spin-wave modes \cite{Harms2024}. Whether an analogous mechanism operates in the present antiferromagnetic crystal, together with the nonlinear saturation of possible unstable modes, remains an important direction for future work. 


\textit{Conclusions and outlook.}---
We have shown that antiferromagnetic spin waves in the presence of an inhomogeneous spin-current-induced flow provide a natural platform for realizing magnonic black-hole and white-hole horizon structures. The two elementary cavity configurations considered here constitute the scattering building blocks of the periodic magnonic crystal introduced in this work. While the BH-WH configuration exhibits resonant transmission satisfying $\mathcal{R}+\mathcal{T} =1$, the complementary WH-BH configuration supports positive and negative-norm transmission channels obeying $\mathcal{T}_{+}-\mathcal{T}_{-} = 1$, revealing a superradiance-like amplification mechanism through mode mixing.

More importantly, a periodic alternation of submagnonic and supermagnonic regions gives rise to an asymmetric magnonic band structure controlled by the spin-current flow. In the purely submagnonic configuration, the relevant bulk gap remains open, whereas entering the supermagnonic regime makes negative-norm propagating channels available and enables the gap to close and reopen, thereby allowing the Zak phase to change. The resulting nontrivial phase hosts a localized hybrid magnon-antimagnon edge state. These results establish a direct connection between analogue-horizon physics and topological magnonics, showing that the horizon structure provides a physical mechanism for electrically controlling magnonic band topology. The resulting non-reciprocal band structure may furthermore enable momentum-selective magnon-antimagnon filtering, providing a possible route toward electrically controlled spin-wave transport in magnonic devices.

S.A. acknowledges funding from Fondecyt Regular 1261323 and ANID CEDENNA CIA 250002. D. G.-P. acknowledges ANID-Subdirección de Capital Humano/Doctorado Nacional/2023-21230818. A.S.N. acknowledges funding from  ANID CEDENNA CIA 250002 and Fondecyt Regular 1230515. 

%


\onecolumngrid
\renewcommand{\thefigure}{S\arabic{figure}}
\renewcommand{\theequation}{S\arabic{equation}}
\renewcommand{\thesection}{S\arabic{section}}
\setcounter{figure}{0}
\setcounter{equation}{0}

\clearpage
\appendix

\begin{center}
    {\Large\bfseries APPENDIX}
\end{center}

\vspace{0.5cm}

\section{Dynamics of AFM}

We consider an antiferromagnetic material described by the following Lagrangian density:

\begin{equation}
    \mathcal{L} = \frac{M_s^2}{2|\gamma|^2 \Lambda} |\dot{\textbf{n}}|^2+ \frac{M_s}{|\gamma| \Lambda} \textbf{H} \cdot (\textbf{n} \times \partial_t \textbf{n}) +  
    \frac{M_s}{2 \Lambda} | \textbf{H} \times \textbf{n}|^2  + \frac{A}{2}|\nabla \textbf{n}|^2 + \frac{K_z}{2} n_z^2 
\end{equation}

Here $\textbf{n}$ denotes the N\'{e}el vector describing the staggered magnetization of the antiferromagnet, while $\textbf{H}$ represents an external magnetic field. The parameter $M_s$ is the saturation magnetization, $\gamma$ is the gyromagnetic ratio, and $\Lambda$ characterizes the strength of the exchange interaction. The constant $A$ corresponds to the exchange stiffnes and $K_z$ is the uniaxial anisotropy parameter that favors alignment of the N\'{e}el vector along the $z$-axis. 

To analyze small excitations around the equilibrium configuration, we consider transverse deviation of the N\'{e}el vector. These fluctuations can be expresed in terms of the complex field, $\psi =  \delta n_x - i \delta n _y $. In terms of this variable, the equation of motion takes the form, for $H=0$:

\begin{equation}
    -K_z \psi + A \frac{\partial ^2}{\partial x^2}\psi - \frac{1}{\Lambda} \left(  \frac{M_s}{|\gamma|}\partial_t \right)^2 \psi =0 
\end{equation}

\section*{Dynamic of AFM with adiabatic STT}

We now consider the effect of the adiabatic spin-transfer torque generated by a spin-polarized charge current. In terms of the net magnetization $\textbf{m}$ and the N\'{e}el vector $\textbf{n}$, the current-induced torques can be written as:
\begin{equation}
    \tau_m = \frac{M_s}{N} \left( \frac{P_m}{e} \textbf{j}_c \cdot \nabla \right)\textbf{m}
\end{equation}
 \begin{equation}
         \tau_n = \frac{1}{N} \left( \frac{P_n}{e} \textbf{j}_c \cdot \nabla \right)\textbf{n}
 \end{equation}

 here $\tau_m$ and $\tau_n$ represents the torque above the net magnetization and the N\'{e}el vector, respectively. The dimensionless parameters $P_m,P_n$ characterize the corresponding spin-transfer efficiencies, N is the density of localized spins, and $\textbf{j}_c$ is the charge current density. 
 
For a one-dimensional system with the current applied along the $x$ direction, these contributions can be expressed as:
 \begin{equation}
     \tau _m =\textbf{u}_m \partial_x \textbf{m} ; \quad \tau_n = \textbf{u}_n \partial_x \textbf{n}
 \end{equation}

 Where $\textbf{u}_m$ and $\textbf{u}_n$ are the corresponding drift velocities.

 In the strong-exchange limit, the net magnetization is a slave variable determined by the dynamics of the N\'{e}el vector. It can therefore be written as

 \begin{equation}
     \textbf{m} = \frac{1}{\Omega_{ex}} (D_n \textbf{n}) \times \textbf{n}
 \end{equation}
 where we have defined $D_n = \partial_t + u_n \partial_x$. The equation governing the net magnetization is
 \begin{equation}
     D_m \textbf{m} = \frac{1}{M_s \hbar} \textbf{n} \times \frac{\delta U}{\delta \textbf{n}}
 \end{equation}

 Where $U$ denotes the potential-energy functional that depends on the N\'{e}el vector in Eq. (1). Assuming that the drift velocities differ only weakly, i.e $u_m = u_n + \delta u$, with $\delta u \ll u_n$, and neglecting corrections proportional to $\delta u$, we finally linearize around the equilibrium configuration $\textbf{n}_o =  \textbf{z}$, we obtain:

 \begin{equation}
     -\frac{1}{\Lambda}\left( \frac{M_s}{|\gamma|} \right)^2 (\partial_t +u \partial_x)^2 \psi +A \partial_x^2 \psi -K_z \psi =0
 \end{equation}

\section*{Conserved current}

Considering eq. (8) for a mode of definite frequency of the form $\psi(x,t) = \phi(x) \text{e}^{-i \omega t}$, the equation of motion reduces to:
\begin{equation}
    (u^2-v^2) \phi'' -2 i \omega u \phi'  + (m^2 - \omega ^2 ) \phi =0 
\end{equation}
To derive a conserved quantity, we multiply this equation by the complex conjugate $\phi^*$, and substract from it the complex conjugate of the entire equation multiplied by $\phi$. After some algebra, this procedure leads to a continuity equation of the form:
\begin{equation}
    \frac{d}{dx} \mathcal{J} =0,
\end{equation}
which defines a conserved current $\mathcal{J}$. The explicit expression for the current is given by:
\begin{equation}
    \mathcal{J} \equiv (u^2-v^2)(\phi^* \phi ' - \phi \phi'^{*}) - 2 i \omega u |\phi |^2   
\end{equation}

For a plane-wave solution of the form $\phi(x) = A\text{e}^{i k x}$, the conserved current simplifies to:
\begin{equation}
    \mathcal{J} = 2 |A|^2 \left( \omega u - (u^2 - v^2 ) k \right)
\end{equation}

\section{Wave-vectors BH-WH cavity }

 In region I, where $u_1<v \to (u_1^2 - v^2)<0$, we can identify two propagating modes, an incident mode and a counter-propagating one: 

     \begin{equation}
        k_\rightarrow =  \frac{ \sqrt{v^2 \omega^2 - m^2 (v^2-u_1^2)}-u_1 \omega}{v^2-u_1^2}   ;
    \end{equation}
     \begin{equation}
        k_\leftarrow = - \left( \frac{u_1 \omega + \sqrt{v^2 \omega^2 - m^2 (v^2-u_1^2)}}{v^2-u_1^2} \right) ;
    \end{equation}

   In region II, where $u_2>v \to (u_2^2 - v^2 ) > 0$, we identify two right propagating modes, as no counter-propagating modes can exist in the supersonic regime:

   \begin{equation}
        k_{-} = \frac{u_2 \omega - \sqrt{v^2 \omega^2 + m^2 (u_2^2 - v^2)}}{(u_2^2-v^2)}
    \end{equation}
     \begin{equation}
        k_{+} = \frac{u_2 \omega + \sqrt{v^2 \omega^2 + m^2 (u_2^2 - v^2)}}{(u_2^2-v^2)}
    \end{equation}

    Finally, in region III, where $u_3< v \to (u_3^2 - v^2)<0$, we obtain results analogous to those of region I. However, in this case, we only consider the right propagating transmitted mode:
        \begin{equation}
        k_3 = \frac{ \sqrt{v^2 \omega^2 - m^2 (v^2-u_3^2)}-u_3 \omega}{v^2-u_3^2}
    \end{equation}

    Considering these, for the first configuration, we propose the following plane-wave solution:

    \begin{equation}
    \phi_I(x) = \text{e}^{ik_\rightarrow x} + R \text{e}^{i k_\leftarrow x } ; \quad \phi_{II}(x) = A\text{e}^{ik_{-} x} + B \text{e}^{i k_+ x } ; 
\end{equation}
\begin{equation}
    \phi_{III}(x) = T\text{e}^{ik_3 x} 
\end{equation}

    In the first cavity, the continuity equations are:  \begin{equation}
       \phi_{I}(0) =  \phi_{II}(0) ;
 \end{equation}
 \begin{equation}
     \alpha_1\phi'_{I}(0)-i \omega u_1 \phi_I(0) = \alpha_2 \phi'_{II}(0) - i \omega u_2 \phi_{II}(0)
 \end{equation}

 and at the interface between region II and III, the continuity conditions are given by:
  \begin{equation}
       \phi_{II}(L) = \phi_{III}(L)
  \end{equation}
  \begin{equation}
       \alpha_2 \phi'_{II}(L)-i \omega u_2 \phi_{II}(L) = \alpha_3 \phi'_{III}(L) - i \omega u_3 \phi_{III}(L)
  \end{equation}

\section{Wave-vectors WH-BH cavity }

   \begin{equation}
        \to k_{1+}= \frac{u_1 \omega + \sqrt{v^2 \omega^2 + m^2 (u_1^2 - v^2)}}{(u_1^2-v^2)}
    \end{equation}

        \begin{equation}
        \to k_{1-} = \frac{u_1 \omega - \sqrt{v^2 \omega^2 + m^2 (u_1^2 - v^2)}}{(u_1^2-v^2)}
    \end{equation}

In region II, where $u_2<v \to (u_2^2-v^2 ) <0$,the system supports both propagating and counter-propagating modes: 

   \begin{equation}
        \to k_{\leftarrow} =- \left( \frac{u_2 \omega + \sqrt{v^2 \omega^2 + m^2 (u_2^2 - v^2)}}{(v^2-u_2^2)} \right)
    \end{equation}

        \begin{equation}
        \to k_{\rightarrow} = \frac{ \sqrt{v^2 \omega^2 + m^2 (u_2^2 - v^2)}-u_2\omega}{(v^2-u_2^2)}
    \end{equation}

In region III, only two right-propagating modes are supported, therefore, two transmitted modes are present:

    \begin{equation}
        \to k_{3+}= \frac{u_3 \omega + \sqrt{v^2 \omega^2 + m^2 (u_3^2 - v^2)}}{(u_3^2-v^2)}
    \end{equation}

        \begin{equation}
        \to k_{3-} = \frac{u_3 \omega - \sqrt{v^2 \omega^2 + m^2 (u_3^2 - v^2)}}{(u_3^2-v^2)}
    \end{equation}

    Based on the above considerations, we propose the following ansatz for each region: 

    \begin{equation}
    \phi_{I}(x) = \text{e}^{i k_1x} ; \quad \phi_{II}(x) = A \text{e}^{ik_{\rightarrow}x} + B \text{e}^{ik_{\leftarrow}x} ; 
\end{equation}
\begin{equation}
    \phi_{III}(x) = T_{+}\text{e}^{i k_{3+}x} + T_{-}\text{e}^{ik_{3-}x}
\end{equation}

\section{Transmission and Reflection coefficients}
For each cavity, we define the reflection and transmission coefficient as the ratios between the reflected/transmitted current and the incident current. For the first cavity, these coefficients are given by:

\begin{equation}
    \mathcal{R} = \frac{|\mathcal{J}_r|}{\mathcal{|J}_{in}|} = \frac{|\alpha_1k_\leftarrow (\omega)-u_1\omega|}{|\alpha_1k_{\rightarrow}(\omega)-u_1 \omega|} |R|^2 
\end{equation}

\begin{equation}
        \mathcal{T} =\frac{|\mathcal{J}_t|}{\mathcal{|J}_{in}|} = \frac{|\alpha_3k_3 (\omega)-u_3\omega|}{|\alpha_1k_{\rightarrow}(\omega)-u_1 \omega|} |T|^2 
\end{equation}
In particular, for the case $u_1 =u_3$ in the BH-WH cavity, the external media are identical. Therefore, from the previous expressions, we obtain:
\begin{equation}
     \frac{|\mathcal{J}_r|}{\mathcal{|J}_{in}|}  = |R|^2= \mathcal{R} ; \quad \frac{|\mathcal{J}_t|}{\mathcal{|J}_{in}|}  = |T|^2 = \mathcal{T} 
\end{equation}
From the current conservation, i.e $\mathcal{J}_{in} = -\mathcal{J}_{r} + \mathcal{J}_{t}$, it follows that the reflection and transmission coefficients satisfy:
\begin{equation}
    \mathcal{R} + \mathcal{T} = 1
\end{equation}

\begin{figure*}[t]
    \centering
    \includegraphics[width=1.0\linewidth]{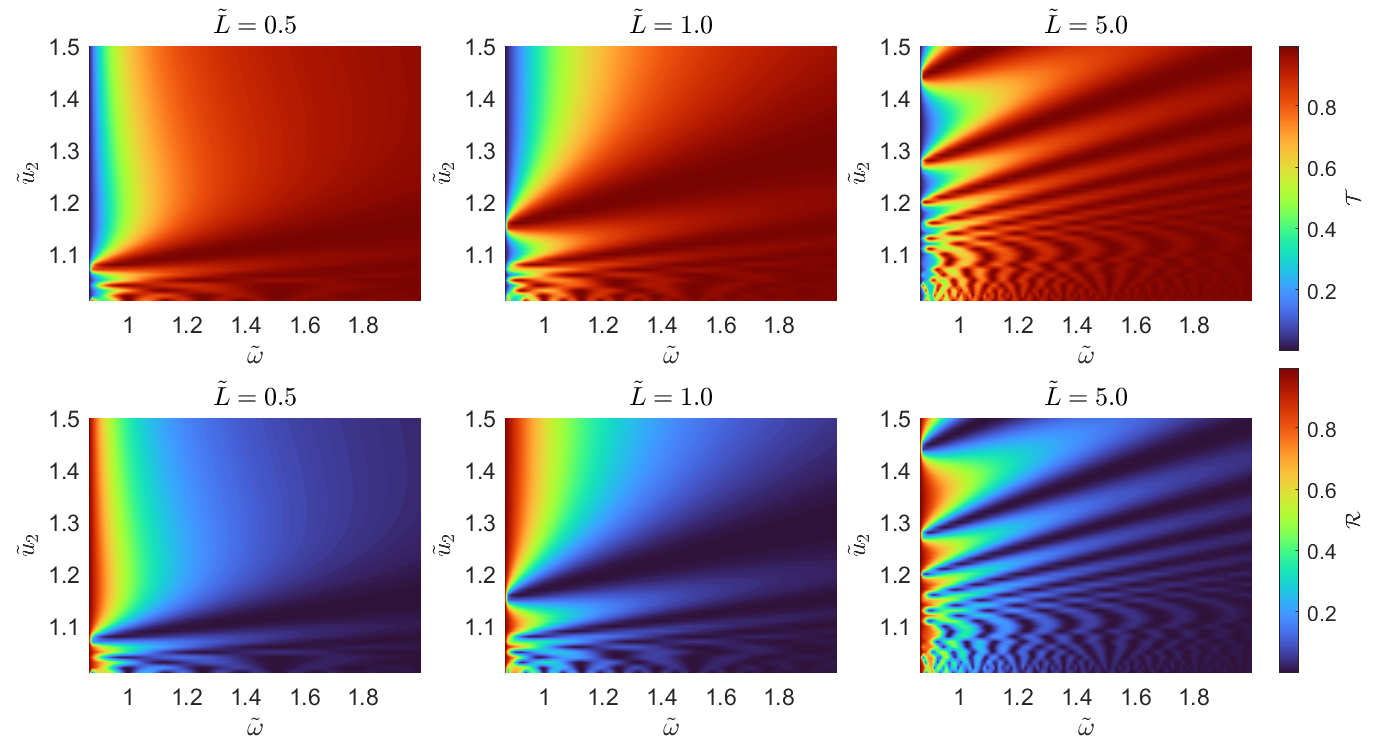}
    \caption{Transmission $\mathcal{T}(\tilde{\omega})$ and reflection $\mathcal{R}(\tilde{\omega})$ coefficients for different cavity lengths $\tilde{L}= L/a$, in function of the supersonic flux current $\tilde{u_2} = u_2/v$ and frequency mode $\tilde{\omega} = \omega/m$, with $\tilde{u}_1 = 0.5$. Top panels correspond to tranmission, while bottom panels show reflection.}
    \label{fig:placeholder}
\end{figure*}

For the WH-BH cavity, there is no reflected wave. Instead, two transmitted channels appear, whose coefficients are given by:

\begin{equation}
        \mathcal{T}_{+} =\frac{|\mathcal{J}_{t+}|}{\mathcal{|J}_{in}|} = \frac{|\alpha_3k_{3+} (\omega)-u_3\omega|}{|\alpha_1k_{\rightarrow}(\omega)-u_1 \omega|} |T_{+}|^2 
\end{equation}

\begin{equation}
        \mathcal{T}_{-} =\frac{|\mathcal{J}_{t-}|}{\mathcal{|J}_{in}|} = \frac{|\alpha_3k_{3-} (\omega)-u_3\omega|}{|\alpha_1k_{\rightarrow}(\omega)-u_1 \omega|} |T_{-}|^2 
\end{equation}

In this case, current conservation, $\mathcal{J}_{in} = \mathcal{J}_{t+} + \mathcal{J}_{t-}$, together with the sign of each current contribution, leads to the condition:
\begin{equation}
    \mathcal{T}_{+}-\mathcal{T}_- = 1
\end{equation}

\section*{Black-White Holes Magnonic Crystal}

We now consider a periodic array of BH-WH-bounded configurations. 
Each unit cell consists of a submagnonic region followed by a supermagnonic region, with lengths \(L_a\) and \(L_b\), respectively. 
Within each region, the field can be written as a superposition of plane-wave solutions,
\begin{equation}
    \phi_1(x) = A\text{e}^{ik_{1+}x} + B\text{e}^{ik_{1-}x}; \quad \quad     \phi_2(x) = C\text{e}^{ik_{2+}x} + D\text{e}^{ik_{2-}x}
\end{equation}

where \(\phi_1(x)\) and \(\phi_2(x)\) denote the solutions in the submagnonic and supermagnonic regions, respectively.

The coefficients of the wave function at the beginning of two adjacent unit cells are related through a transfer matrix,

\begin{equation}
    \mathbf{A}_{n+1}=\mathcal{M}(\omega)\mathbf{A}_{n}
\end{equation}

where \(\mathbf{A}_{n}=(A_n,B_n)^T\), and \(\mathcal{M}(\omega)\) is the transfer matrix of one unit cell. 
In terms of the matching matrices at the interfaces and the propagation matrices inside each region, this matrix is given by

\begin{equation}
    \mathcal{M}(\omega) = \mathcal{M}_4^{-1} \mathcal{M}_3 \mathcal{M}_2^{-1} \mathcal{M}_1
\end{equation}

The auxiliary matrices are defined as

\[
M_1 =
\begin{pmatrix}
e^{i k_r L_a} & e^{i k_l L_a} \\
(a_1 k_r - u_1 \omega)e^{i k_r L_a} &
(a_1 k_l - u_1 \omega)e^{i k_l L_a}
\end{pmatrix},
\]

\[
M_2 =
\begin{pmatrix}
1 & 1 \\
a_2 k_{2m} - u_2 \omega &
a_2 k_{2p} - u_2 \omega
\end{pmatrix},
\]

\[
M_3 =
\begin{pmatrix}
e^{i k_{2m} L_b} & e^{i k_{2p} L_b} \\
(a_2 k_{2m} - u_2 \omega)e^{i k_{2m} L_b} &
(a_2 k_{2p} - u_2 \omega)e^{i k_{2p} L_b}
\end{pmatrix},
\]

\[
M_4 =
\begin{pmatrix}
1 & 1 \\
a_1 k_r - u_1 \omega &
a_1 k_l - u_1 \omega
\end{pmatrix}.
\]

For an infinite periodic system, Bloch's theorem imposes

\begin{equation*}
    \mathbf{A}_{n+1}=\mathcal{M}(\omega)\mathbf{A}_{n} = e^{iqd}\mathbf{A}_{n}
\end{equation*}
where \(d=L_a+L_b\) is the lattice period and \(q\) is the Bloch wave number. 
Therefore, the allowed bands are obtained by solving the eigenvalue problem

\begin{equation}
    \det\!\left[\mathcal{M}(\omega)-e^{i\tilde q d}\mathbb{I}\right]=0.
\end{equation}

An alternative and complementary way to understand the band asymmetry is obtained by applying a local gauge transformation to the field in each region,

\begin{equation}
    \phi_j = \text{e}^{i \Theta_j x} \chi ; \qquad \chi = C \text{e}^{iq_j x}+ D \text{e}^{-iq_j x}
\end{equation}

where,
\begin{equation}
    \Theta_j (\omega) = \frac{u_j \omega}{u_j^2-v^2}  ; \qquad q_j  = \frac{\sqrt{v^2 \omega^2 +m^2(u_j^2-v^2)}}{u_j^2-v^2}
\end{equation}

With this transformation, the equation of motion in each region reduces to the symmetric Helmholtz-like form
\begin{equation}
    \chi'' + q_j^2 \chi =0
\end{equation}

In this gauge-transformed frame, the Bloch condition reads

\begin{equation}
    \chi(x+d) = \text{e}^{i \tilde{K}d}\chi (x)
\end{equation}

and the resulting band structure is symmetric in the quasi-momentum \(\tilde{K}\). 
However, the original field \(\phi_j\) contains an additional phase accumulated across the unit cell. 
As a consequence, the Bloch wave numbers in the two descriptions are related by

\begin{equation}
    \tilde{K} = \tilde{q}-\Gamma (\omega) ; \qquad  \Gamma(\omega) = \frac{ \Theta_1 L_A + \Theta_2L_B}{d}
\end{equation}

Therefore, the inhomogeneous flow induces a frequency-dependent momentum shift, linear in \(\omega\), which is responsible for the non-reciprocal and asymmetric band structure of the original system.

\section*{Topology of the Crystal}
To understand the topology of the system, we can decompose the unit-cell matrix $\mathcal{M}(\omega)$ as:
\begin{equation}
    \mathcal{M}(\omega) = P D_2 P^-1 D_1
\end{equation}
where $P = M_4^{-1} M_2$ and $D_{1,2}$ are diagonal matrices:
\begin{equation}
    D_1 = 
\begin{pmatrix}
e^{i k_r L_a} & 0 \\
0 &
e^{i k_l L_a}
\end{pmatrix} ; \qquad 
D_2 = 
\begin{pmatrix}
e^{i k_{2-} L_b} & 0 \\
0 &
e^{i k_{2+} L_b}
\end{pmatrix}
\end{equation}

Taking the determinant, we obtain:
\begin{equation}
    det(M) = det(P) \cdot det(D_2) \cdot det(P^{-1}) \cdot det(D_1) = det(D_2) \cdot det(D_1)
\end{equation}
\begin{equation*}
    \to det(M) = e^{i L_a (k_r + k_l) + i L_b (k_{2-}+k_{2+})}
\end{equation*}
Using the wave-vector of each region :
\begin{equation}
    k_r + k_l = \frac{2 u_1 \omega}{\alpha_1} ;  \quad k_{2-} + k_{2+} = \frac{2 u_2  \omega}{\alpha_2} 
\end{equation}
We finally obtain:
\begin{equation}
    det(M) = e^{i \theta (\omega)} ; \quad \theta(\omega) = 2(\frac{u_1 L_a}{\alpha_1} + \frac{u_2 L_b}{\alpha_2}) \omega
\end{equation}

With this in mind, we define the auxiliary matrix $\tilde{M} = e^{-i \theta(\omega)/2} M$, with:
\begin{equation}
    det(\tilde{M} ) = det(e^{-i \theta /2} \mathbb{I} ) det(M) = 1
\end{equation}

After some algebra, the auxiliary matrix $\tilde{M}$ takes the form:

\begin{equation}
    \tilde{M} = \begin{pmatrix}
        \alpha & \beta \\
        \beta^* & \alpha^*
    \end{pmatrix}
\end{equation}
Since the determinant is unity:
\begin{equation}
    |\alpha|^2 - |\beta|^2 = 1
\end{equation}
Therefore, $\tilde{M}$ belongs to the group SU(1,1). Matrices in this group satisfy $\tilde{M} ^*  = \sigma_x \tilde{M} \sigma_x$. Consequently, if $u_q$ is an eigenvector of $\tilde{M}$ related with the eigenvalue $\lambda = e^{iqd}$, then:
\begin{equation}
    u_{-q} = \sigma_x u_q ^* 
\end{equation}
Since $M$ and $\tilde{M}$ differ only by a phase factor, they share the same eigenvectors. The corresponding Berry conection:
\begin{equation}
    A(q) = i\langle u_q | \partial_q u_q \rangle = i u_q^{\dagger} \sigma_z\partial_q u_q
\end{equation}
Using the above relation between the eigenvectors, one finds that $A(-q) = -A(q)$. Combined with inversion symmetry, this relation leads to the quantization of the Zak phase.

\section{Closing gap for pure submagnonic crystal}

In the symmetric case, which as shown previously, preserves the same band-gap condition as the original system, the band-gap structure is determined by:
\begin{equation}
    cos(qd) = cos(k_1 L_a ) cos(k_2L_b) + sin(k_1 L_a ) sin (k_2 L_b) f
\end{equation}
where we have defined $f = \frac{\alpha_2^2 k_2^2 + \alpha_1^2 k_1^2}{2 \alpha_1 \alpha_2 k_1 k_2} = - C$. Defining $\delta_j = k_j L_j $, the dispersion relation becomes:
\begin{equation}
    F(\omega ) = \cos(qd) = \cos(\delta_1) \cos(\delta_2 ) - C \sin(\delta_1) \sin(\delta_2 ) 
\end{equation}
For the case $u_1,u_2<v$, the band-gap closing (Bragg) condition found previously is:
\begin{equation}
    |k_1| L_a + |k_2| L_b = n\pi \to \delta_1 + \delta_2 = \pi
\end{equation}

After some algebra, we obtain:
\begin{equation}
    F = -1 - (C-1) sin^2 (\delta_1)
\end{equation}

Since $C>  = 1 $ in the purely submagnonic regime, the gap-closing conditions, $F (\tilde{q}) =\pm 1 $ and $F'(\tilde{q}) = 0$, cannot be satisfied, since $\sin (\delta_1) \neq 0$ for construction. Therefore, no gap closing occurs in the purely submagnonic case, except the trivial configuration $u_1 = u_2$

\section*{Edge-mode nature}

The Bloch problem in the SU(1,1) representation takes de form:
\begin{equation}
    \tilde{\mathcal{M}}u_{\tilde{q}} = \tilde{\lambda}u_{\tilde{q}} 
\end{equation}
where $\tilde{\lambda} = e^{i \tilde{q}d}$. In the gap, the condition over the system is $|\text{Tr}(\tilde{\mathcal{M}})| >2$, with $|\tilde{\lambda}| \neq 1$. Therefore, it can be proved that:
\begin{equation}
    (|\tilde{\lambda}|^2 -1 ) u_{\tilde{q}}^{\dagger} \sigma_z u_{\tilde{q}} =0 \quad \to \quad   u_{\tilde{q}}^{\dagger} \sigma_z u_{\tilde{q}} = 0
\end{equation}

Therefore, the eigenvector in the band-gap must be an isotropic Klein vector, i.e, it must contain both positive and negative-norm components. In our system, this implies that the edge mode must be an hybrid magnon-antimagnon state.

\end{document}